Nanomaterials Science: Perspective

# The common attribute shared by defects, surfaces, and nanostructures: the BOLS-NEP notion

Chang Q Sun[1,2]

*Atomic undercoordination fascinates defects, surfaces, and nanostructures in electronic binding energy, lattice oscillation frequency, elasticity and plasticity (IHPR), thermal stability, photon emisibility, reactivity, dielectrics, super-hydrophobicity, spin-resolved topological edge and monolayer high-$T_C$ superconductivity, etc., through local bond contraction, quantum entrapment and polarization.*

Contents

Highlight

- *Atomic undercoordination matters substance through bond contraction and electron occupancy*
- *Nanocrystal prefers a core-shelled structure with a skin of two atomic-layer-spacing thick*
- *Atomic cohesive energy dictates the thermal stability and the critical energy for phase transition*
- *Competition between energy density and cohesive energy determines the IHPR strongest size*
- *Spin-resolved polarization fosters the edge topological insulator and high-$T_c$ superconductivity*

**Abstract**

This article describes a consistent insight into the unusual performance of adatoms, point defects, terrace edges, surfaces, and nanostructures of various shapes from the perspective of atomic undercoordination. The notion of bond order-length-strength correlation and the associated nonbonding electron polarization (BOLS-NEP) features that bond between undercoordinated atoms contracts spontaneously associated with its strength gain. Bond contraction raises the local density of charge and energy and the bond strength gain deepens the interatomic potential well to trap the core and bonding electrons. In turn, the locally and densely entrapped binding electrons polarize those in the valence band and above pertained to the even-less coordinated atoms at the terminal edges. The BOLS-NEP notion reconciles the unusual behaviors of undercoordinated systems and the size matter of nanostructures on lattice oscillating dynamics, mechanical strength, thermal stability, photon emisibility, chemical reactivity, dielectric permeability, edge Dirac-Fermion and monolayer high-$T_C$ superconductivity, etc.

# 1    Wonders of atomic undercoordination

As an independent degree of freedom, atomic undercoordination has fascinated enormously adatoms, point defects, surfaces, terrace edges, and nanostructures of various shapes for academic interest and applications in industrial sectors, which has been recognized as one of the major economic, scientific, social, and technological thrusts in the current century. Atoms at edges of a surface are the key components that host the charge carriers of topological insulators and high-$T_C$ superconductors. The high reactivity of terrace edges has boomed the single atom catalysis. As the crystal size reduces, all detectable qualities remain no longer the bulk constant but they vary with the shape and size, or the fraction of the undercoordinated atoms of the entire specimen. Concerned properties include elastic modulus, critical energy of phase transition, vibrational phonon frequencies, electronic binding energies, semiconductor bandgap, dielectric constant, and many more.

With the reduction of crystal size at the sub-micrometer scale, the yield strength of a material increases with the inverse of the square-root of its feature size, known as the Hall-Petch relationship (HPR). At the $10^{0-1}$ nm scale, the HPR reverts its strength maximum with size and a strongest size presents around 10 nm size, called inverse HPR [1, 2]. The IHPR is 2-5 times as strong as the bulk, making an artificially twinned diamond and a boron nitride nanocrystal at the nanometer scale the hardest substance ever known [3, 4]. Further size reduction will fosters the super plasticity occurring to monoatomic chain and nanowires. Stretching elongates the Au-Au distance from its $0.23 \pm 0.02$ nm at 4 K to 0.48 nm at room temperature, while it is 0.288 nm in the bulk [5]. The catalytic yield can reach orders high as the reactant size is reduced to the monomer or dimer, nurtured the concept of single atom catalyst.

Despite the fascination of the undercoordinated systems and the booming of engineering and sciences regarding defect, surface, and nanostructure, progress in theoretical description and consistent understanding remain infancy. Generally, one phenomenon is often associated with numerous debating theories from various perspectives such as the size-reduction induced blueshift of semiconductor photo-luminesce of semiconductors.

In fact, the relaxation of bond length and energy and the manner of electron occupancy determine the detectable properties of a substance through perturbing the Hamiltonian according the solid state quantum theory. This presentation describes a consistent understanding of the undercoordinated systems from the perspective of atomic undercoordination induced bond contraction and quantum entrapment and polarization (BOLS-NEP) [5]. Consistency between theoretical predictions demonstrated the impact and profoundness of the BOLS-NEP theory to reconciling the performance of the undercoordinated systems.

## 2 BOLS-NEP theory
### 2.1 BOLS-NEP notion

Except for the zero-coordinated ($z_0 = 0$), or an isolated atom and the fully-coordinated ($z_b = 12$) atom in the referencial bulk interior of a fcc crystal, all rest in the universe are undercoordinated ones such as adatoms and those presenting at sites of surfeces, point defects, terrace edges, and the inner skin of a cavity. The cavity forms the basic element of porous foams, and the metal-organic framewok (MOFs) that has been widely used for disalnation of water harvesting. Nanostructures of various dimensions and dimensionalties, including one-dimensional atomic chains, and two-dimensional atomic sheets have high fraction of undercoordinated atoms. A three-dimensional nanostructure is a collection of high-fraction of undercoordinated atoms in the curved skin, for instance. For an atom having a $z_b$ neighbours in the bulk, its effective atomic coordination number (*CN*) is reduced to $z = 12 \times z_c/z_{cb}$ for the universility of the theory, such as the *bcc* structure $z_{cb} = 8$. For the diamond structure, the atomc *CN* is 12 rather than 4 because of the diamond is an interlock of two fcc structures.

Atomic CN is an important degree of freedoms that has nurtured the sciences of surface, defect, and nanostructures. One can imagine, what will happen to the universe without atomic undercoordination. Atoms having different types of atomic neibours, known as heterocoordination, which is another issue of coencern [6]. Performance of electrons and chemical bonds of an undercoordinated system follow the bond order-length-strnegth correlation and nonbonding electron polarization (BOLS-NEP) notion, as illustrated in Figure 1 and formulated as follows [5]:

$$\begin{cases} C_z = d_z / d_b = 2\{1 + \exp[(12-z)/(8z)]\}^{-1} & (bond-contraction-coefficient) \\ C_z^{-m} = E_z / E_b & (bond-strengthening-coefficient) \end{cases}$$
(1)

where z is the atomic *CN* and *m* is the bond nature index to be optimized in practice. For metals, $m = 1$, for carbond and silison, $m = 2.56$ and $4.88$. When the atomic *CN* is reduced from the bulk standard to z, the bond length and energy ($d_b$, $E_b$) transit to ($d_z$, $E_z$) sponteneously.

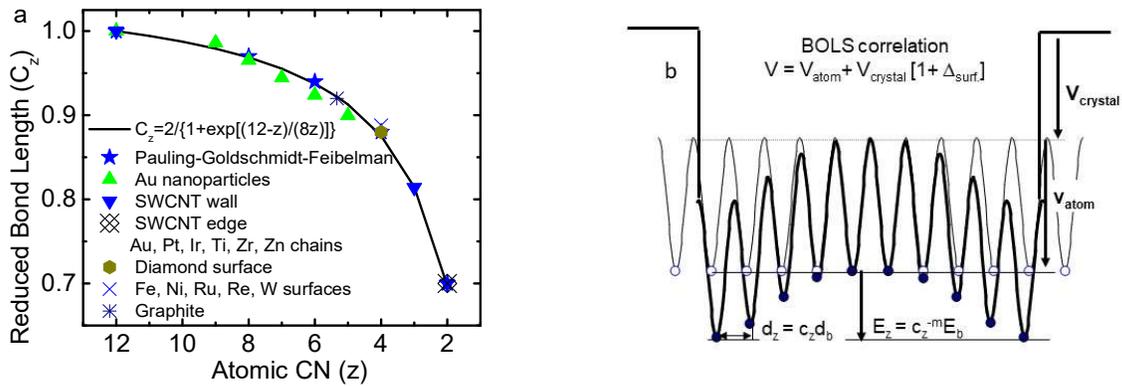

Figure 1 BOLS-NEP notion for the atomic undercoordination induced (a) bond contraction and (b) potential well depression at site nearby atomic vacancies or bonding network terminals [5].

As the consequence of sponteneous bond contraction, the bond gains its strength. Bond contraction raises the local charge and enenrgy enesity. Bond strength gain deepens the inter-atomic potential well and entraps electrons accomonated in core levels and bonding orbitals. In turn, the locally and deeply entraped core and bonding electrons polarize the nonbonding electrons of the fewer coordinated edge atoms. The polarization raises the upper edge of the valence band to close or suppass the Fermi level.

However, the polarization is subject to the configuration of the nonbonding electrons of the less-coordinated edge atoms. Evidence revealed that polarization ocurs only to those elements having halfly-occupied $s$ orbitals such as Rh, Ag, Au but not those without such electronic occupancy such as Pt and Co atoms. Atomic undercoordination shortens and strengthes the H-O covalent bond but does the O:H nonbond contrastingly for the hydrogen bond (O:H-O with ':' being the elelctron lone pair of oxygen) in water, ice and aqueous solutions [7, 8]. In other words, atomic undercoordination shortens and stiffens bonds between under-coordinated atoms with an assocation of core electron entrapment and subjective nonbonding elelctron polarization, without any exception. The BOLS-NEP provides a perturbation to the crystal potential of the Hamiltonian in the Schödinger eqation that not only features the performance of bonds and elelctrons in the eneregtic-spatial-temporatal domains but also nurtures directly the structure, morphology, and detectable quantities of a substance [5].

2.2    Site and crystal size resolution

Given a specimen of $K$ dimensionless size and $\tau$ dimensionality, its detectable quantity $Q(K)$ varies with respect to its bulk standard $Q(\infty)$[5]:

$$\frac{\Delta Q(\tau,K)}{Q(\infty)} = \begin{cases} \Delta q_z/q & \text{(atomic site resolved)} \\ \sum_{j=0}^{3}\gamma_j(\Delta q_j/q) & \text{(crystal size-and-shape resolved)} \end{cases}$$

$$\gamma_j = dLn[V(\tau,K)] = \tau c_j/K$$

$$z_1 = 4(1-0.75/K), z_2 = z_1+2, z_3 = z_2+4 \text{ (for spherical structure)}$$

(2)

$K$ is the number of atoms lined along the radius of a spherical or a cylindrical shaped crystal. $\tau = 1, 2, 3$ is the dimensionality of a monoatomic chain or a thin slab, a rod and a sphere, respectively. For a flat surface, $1/K = 0$, for a solid and a hollow sphere, $1/K > 0$ and $1/K < 0$, respectively. The sum starts from the outermost atomic layer inward up to a maximum three as the atomic-CN reduction becomes negligible. For the monatomic chains, atomic sheets, adatoms, point defects, and monolayer skins, no sublayer submission is necessary.

The $q$ features the site resolved density of property that depends functionally on the local bond length, energy and charge distribution, which differentiates intrinsically the properties of the undercoordinated atomic site from those in the ideal bulk. If $\Delta q_z = q_z - q = 0$, nothing will happen to defects, surfaces, or nanocrystals. The $\gamma_j$ is the volume ratio of the $j$th atomic layer to the entire body of the nanocrystal. Therefore, the size dependency arises from two parts: one is the intrinsic origin $\Delta q_z \neq 0$ and the other is extrinsic quantity $\gamma_j$, or the fraction of the undercoordinated atoms.

The presently concerned properties in this perspective include:

$$q \propto \begin{cases} E & \text{(band gap } E_g, \text{XPS core level shift } E_v) \\ E^{1/2}/d & \text{(phonon frequency shift } \Delta\omega, \text{bond stiffness)} \\ zE & \text{(cohesive enenrgy } E_{coh} \text{ related to the } T_C \text{ for phase transiton)} \\ E/d^3 & \text{(enenrgy density } E_{den} \text{ related to elasticity } B, \text{ yield strength } \sigma) \end{cases}$$

(3)

For instances, the bond energy determines the energy gap $E_g$ between the valence band and the conduction band of a semiconductor [5]. The static dielectric constant depends inversely with the square of the $E_g$. The binding energy shift of electrons in the core band depends on the bond energy as well according to the tight-binding approximation. The atomic cohesive energy determines the critical temperature $T_C$ for phase transition, and the energy density determines the elasticity at equilibrium (r = d) and the yield strength at plastic deformation (r > d)[5].

3     Computational and spectrometric justification

Diffraction crystallography, scanning microscopy, and electron and phonon spectroscopy features the performance of atoms and electrons in the energy-spatial-temporal domains. The information of bond length and energy and electron distribution nurtures the detectable quantities of a substance. Spectrometrics extends the spectroscopy with focus more on deriving quantitative information regarding the response of bond length, bond energy, and electron occupancy to stimulus by analyzing the spectral signatures.

The mathematical foundation for the spectrometrics is the Fourier transformation that gathers information of bonds vibrating in the same frequency, or electrons of the same binding energy, into a specific peak regardless of their sites in real space, crystal geometry, or structure phase. The peak intensity represents the maximal population and the peak width the fluctuation. The integral of the peak corresponds to the abundance of electrons and phonons collected under certain conditions.

The physical principle of the spectrometrics is the electron binding energy or bond vibrating frequency shift arises form modulating the Hamiltonian through stimulus such as atomic undercoordination, mechanical activation, thermal excitation, charge injection by solvation or chemical doping and absorption. The first two expressions in eq (3) formed the foundations to the XPS electron binding energy shift from the respective level of an isolated atom and the Raman phonon frequency shift from the referential dimer to the states of crystals being perturbed by applied stimulus.

   3.1    Skin bond contraction

Figure 2a shows the inhomogeneous Au-Au bond contraction occurring to the outermost two atomic layers of a 3.8 nm sized crystal. Molecular dynamics (MD) calculations and coherent electron diffraction [9] revealed that the relaxation occurs mainly to the out-of-plane bond contractions for the edge atoms (~0.02 nm, ~7%); a significant contraction (~0.013 nm, 4.5%) for the (100) surface atoms; and a much smaller contraction (~0.005 nm, 2%) for atoms in the (111) facets. Extended X-ray absorption fine structure (EXAFS) measurements [10] revealed the same trends of atomic-*CN* dependence of Au-Au bond contraction that is independent of the type of substrate support. Figure 2b shows the typical mean lattice contraction for Ag and Au thin films formulated using the core-shelled structural pattern given in eq (2). Figure 2 inset b expresses the size dependency of lattice contraction. For the 5.0 nm Ag crystal, 60% of the atoms have the bulk distance but 40% have shorter atomic distances [11]. The average atomic distance for Ag, Cu, and Ni shortens by 1.6 – 2.0% for small crystals and about 0.6% for relatively larger ones. The bond contraction raises the local density of bonding electrons, energy density, and mass. The atomic-site and crystal-size resolved bond length contraction confirmed the BOLS expectation of the core-shell nanostructures.

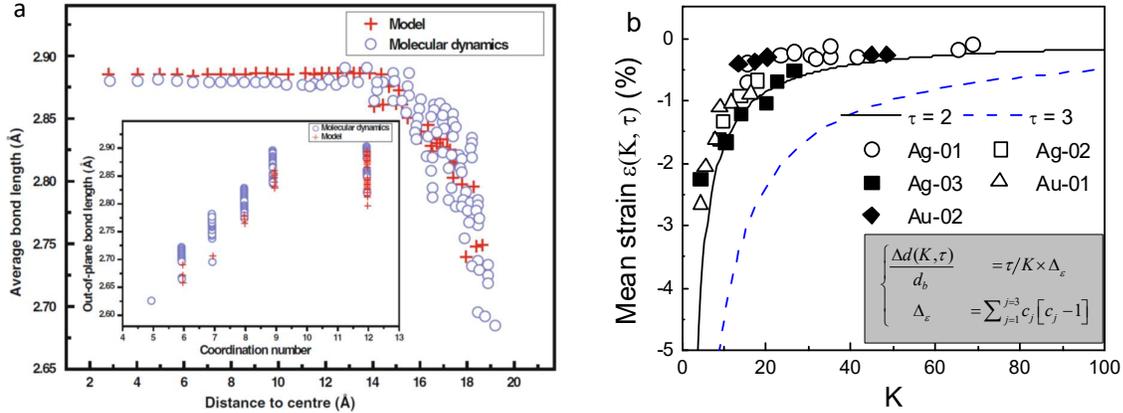

Figure 2  Atomic-site and cluster-size resolved bond contraction at (a) the outermost two atomic shells of a gold nanocluster [9] and (b) BOLS theory reproduced (scattered datum) size dependence of mean lattice contraction of (a) Ag-01 [12], Ag-02 [11], Ag-03 [13], Au-01 [14], and Au-02 [15] thin films ($\tau = 1$).

### 3.2  Skin thickness of the core-shelled structure

The differential phonon spectrometrics (DPS) has enabled determination of the skin thickness of the core-shelled nanocrystals [16]. The DPS is able to distill the number of phonons from their bulk mixture by subtracting the phonon spectral peaks of the sized samples by the referential bulk spectral peak upon all of them being area normalized. The DPS not only determines the skin-shell thickness but also distinguishes the performance of bonds and electrons in the skin shells in terms of length and stiffness.

Figure 3 insets show the peak-area normalized Raman spectra collected from the sized $CeO_2$ nanocrystals under the ambient conditions [17]. The measurements were focused on the vibration longitudinal optical (LO) mode centered at 464 cm$^{-1}$. The peak area normalization to one unit aims to minimizing the experimental artifacts.

The DPS resolves the number of phonons transiting their population from the bulk (valley) to the skin (peak). The DPS blueshift from the valley to the peak represents the skin bond stiffness gain and the redshift is associated with the undercoordination-induced subjective polarization of the surface electrons that screens and splits the local crystal potentials [6]. The CeO$_2$ skin covering sheets show both redshift and blueshift dominated by the bond contraction and the electron polarization.

An integration of the DPS peak corresponds to the number/volume fraction of bonds transiting from the core to the skin covering shell of the nanostructures. For a spherical structure of D across, the $f(D) = \Delta V/V = 3\Delta R/R = 6\Delta R/D$, which gives rise to the shell thickness $\Delta R$ of 0.5 nm for the CeO$_2$ nanocrystals. The 0.5 nm is two atomic diameters. The direct measure of the skin-shell thickness proves for the universal core-shelled nanostructures. The unusual performance of bonds and electrons in the skin-shells and the varied skin/volume ratio dictate the size dependency of nanostructures [5].

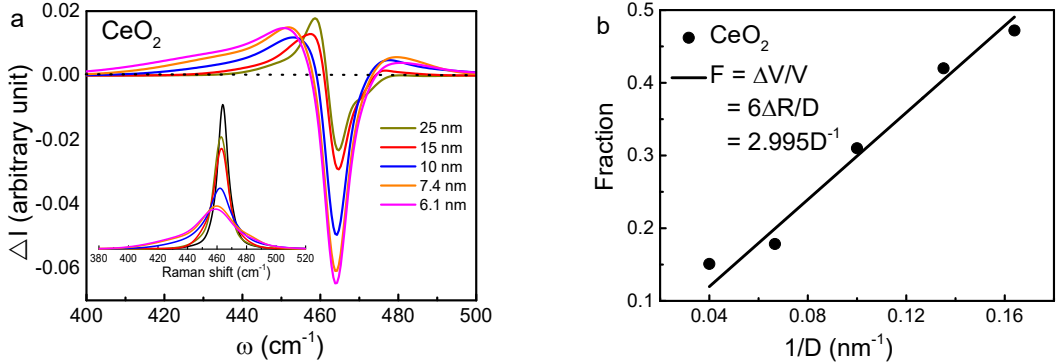

Figure 3. DPS determination of the skin-shell thickness of the core-shelled structured CeO$_2$ nanocrystals [17]. The (a) DPS profiles and (d) their peak integrals for the sized crystals. Inset a shows the peak-area normalized Raman band. The fraction coefficient $f(D) = \Delta V/V = 6\Delta R/D$ confirms the skin-shell thickness of $\Delta R = 0.5$ nm for CeO$_2$ nanocrystals [16].

### 3.3 Phonon frequency shift: dimer and collective vibration

The DPS in Figure 4a resolves the phonon frequency relaxation of the few-layered MoS$_2$ [18]. Phonons transit from the bulk component to the undercoordinated edge and skin component. The collective oscillation of bonds between a certain atom and its nearest z neighbors governs the $A_{1g}$ LO mode frequency redshift while the dimer vibration drives the $E^1_{2g}$ mode blueshift, which are consistent to the D and 2D mode redshift and the G mode blueshift of graphene, respectively [16]. Based on the core-shell structure configuration, one can reproduce the Raman shift of nanocrystals, as shown in Figure 4b, to confirm the local bond length, energy relaxation and determine the bond nature index and the referential dimer vibration frequency $\omega(1)$ from which the Raman shift proceeds, as Table 1 shows.

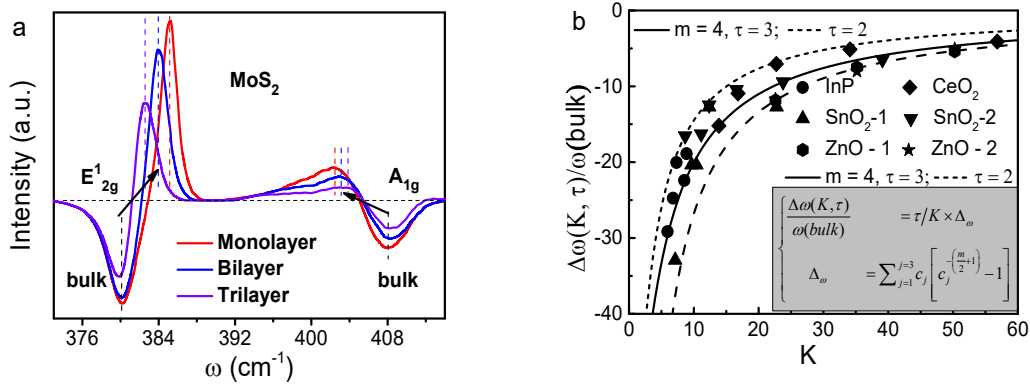

Figure 4. (a) Number-of-layer resolved Raman DPS shift for (a) $MoS_2$ [19, 20] and (b) BOLS theoretical reproduction of the LO mode mean frequency shift for (b) InP [21], $CeO_2$ [17, 22], $SnO_2$ [23], and ZnO [5, 24] nanocrystals. Table 1 shows the $\omega(\infty)$ and the referential $\omega(1)$ derived from simulation.

Table 1. Nanograin dimer vibration frequencies and their redshift with size increases.

| Material | m | $\omega(\infty)(cm^{-1})$ | $\omega(1)$ $(cm^{-1})$ |
|---|---|---|---|
| $CeO_2$ | | 464.5 | 415.1 |
| $SnO_2$ | 4.0 | 637.5 | 613.8 |
| InP | | 347 | 333.5 |
| ZnO | | 441.5 | 380 |

3.3 Electronic binding energy shift: entrapment and polarization

Scanning tunneling microscopy/spectroscopy (STM/S) [25-27], density functional theory (DFT) calculations, and XPS measurements [6] confirmed the core-level quantum entrapment and defect-induced nonbonding electron polarization. Figure 5a shows that bright protrusions surround the point defect on a graphite HOPG(0001) surface. Electrons of atoms nearby the vacancy form a resonant peak at the $E_F$. The spectral peak and protrusion are the same to those appeared at the zigzag edges of monolayer graphene nanoribbons (GNR) [28-30]. The resonant current, known as Dirac-Fermion (DF), flows between the STM tip and the defect edge without bias being applied. However, such resonant states are absent from the flat skin, or the armchair edges. These observations suggest that the point defects and the GNR zigzag-edge share the same identity of bond relaxation, core electron entrapment, and nonbonding electron polarization [6]. The flat surface and the armchair-edge undergo bond contraction but not polarization because of lacking of the local dangling bond electrons.

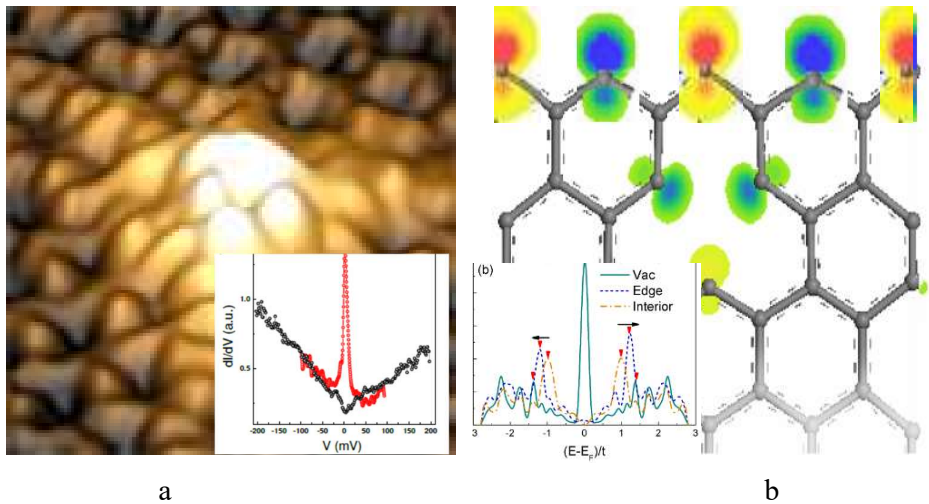

a  b

Figure 5. (a) STM/S protrusions and resonant current probed from the defected HOPG(0001) surface [25] and (b) DFT derived local, spin-resolved density-of-state pertained to GNR zigzag-edge and an atomic vacancy of the monolayer GNR [6].

DFT calculations [6] revealed that the spin-resolved DFs create preferably at zigzag-edge of a GNR or surround an atomic vacancy. The densely entrapped core electrons polarize the dangling σ-bond electrons of atoms of identical √3d distance along the edge, see Figure 5b. DFT derived the same sharp edge states (inset) for atomic vacancy and GNR edge to that of the STS [25]. The locally and densely entrapped bonding electrons pin the DFs through polarization. However, along the armchair-GNR edge and the reconstructed-zigzag-GNR edge, the quasi-triple-bond formation between the nearest edge atoms of *d* distance or less prevents the DFs formation. The presence and absence of the DFs demonstrate the subjective polarization of electrons associated with the even-less coordinated edge atoms.

The XPS C 1s energy shift of the monolayer skin and point defects of the HOPG(0001) surface, and the layered GNR flakes [31], shown in Figure 6, further confirmed the core-level entrapment and nonbonding electron polarization. The skin ZPS differentiates two spectra collected at 75° and at 25°. The defect ZPS differentiates two spectra collected at 50° from the surface after and before high-density defect generation by $Ar^+$ spraying. The color zones of the structures shown by inset a contributing to the excessive ZPS states in each case. The atomic *CN* for the skin is derived about 3.1, which is close to the ideal case of 3.0 of graphene interior. The atomic *CN* for the vacancy extends from 2.2 to 2.4, which indicates that the next nearest neighbors contribute to the ZPS identity of the vacancy.

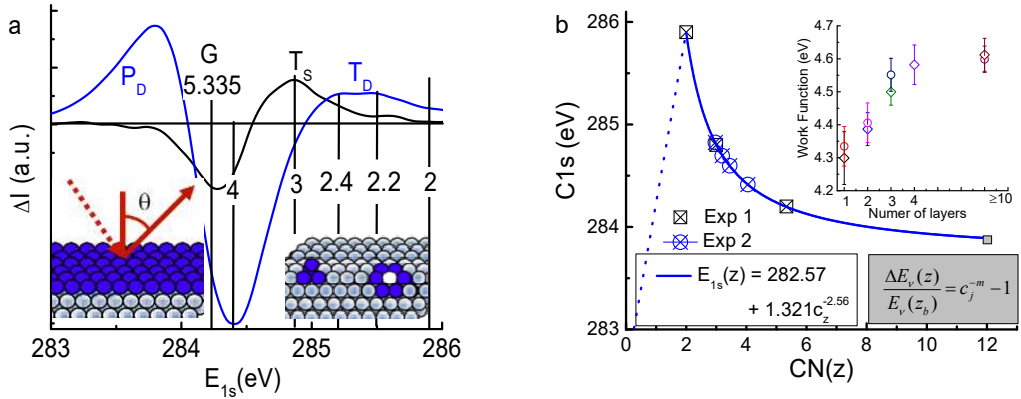

Figure 6. C 1s energy shift for the (a) monolayer skin ($T_S$) and point defects ($T_D$) [6] of HOPG (0001) surface and the (b) atomic-CN resolved C 1s energy for graphene [6] confirm the BOLS expected quantum entrapment. The (a) defect states ($P_D$) and (b) work function reduction (inset b) [31, 32] demonstrate the BOLS-NEP prediction of polarization [6].

The bulk valley G centered at 284.20 eV (z = 5.335) and the 284.40 eV valley is a mixture of the bulk and the skin. The skin $T_S$ (z ~ 3.1) and the defect $T_D$ (z ~ 2.2 ~ 2.4) components denote state transition from the bulk to the entrapment. The $P_D$ component at the upper edge arises from the screening and splitting of the crystal potential by the DFPs [6]. The reduction of work function [32] results from polarization of the dangling bond electrons. These observations follow the same trend demonstrated by $C_{60}$ deposited on CuPc substrate as detected using UPS, XPS, and synchrotron radiation spectroscopy [33].

Strikingly, the *CN*s of atoms annexed the vacancy are compatible to that of the GNR edge of 2.0. One can evaluate the length and strength of the C-C bonds and the C 1*s* shift associated with the undercoordinated atoms, as featured in Table 2. Consistency in the expected effective *CN* and the specific energy evidences sufficiently the accuracy and reliability of the BOLS-NEP derivatives. Most strikingly, only one neighbor loss makes a great difference between C atoms at edges and C atoms in the monolayer skin. The defected P states of C are the same to Rh, Au, Ag, Cu and W adatoms or terrace edges and the skin entrapment is the same to Pt, Re, and Co adatoms or nanocrystals.

Table 2. BOLS-ZPS resolved z-dependent C-C bond length $d_z$, bond energy $E_z$, and C 1s energy of carbon allotropes [6].

| | z | $C_z$ | $d_z$(nm) | $E_z$(eV) | C 1s (eV) | Refs | P (eV) |
|---|---|---|---|---|---|---|---|
| Atom | 0 | - | - | - | 282.57 | - | |
| GNR edge | 2.00 | 0.70 | 0.107 | 1.548 | 285.89 | 285.97 [32] | |
| Graphite vacancy | 2.20 | 0.73 | 0.112 | 1.383 | 285.54 | | 283.85 |
| | 2.40 | 0.76 | 0.116 | 1.262 | 285.28 | - | |
| GNR interior | 3.00 | 0.81 | 0.125 | 1.039 | 284.80 | 284.80 [32]; 284.42 [34]; 284.90 [35]; 284.53-284.74 [36] | |
| Graphite skin | 3.10 | 0.82 | 0.127 | 1.014 | 284.75 | - | |
| Graphite | 5.335 | 0.92 | 0.142 | 0.757 | 284.20 | 284.20 [32]; 284.30;[34, 35]; 284.35 [37]; 284.45 [38] | |
| Diamond | 12.00 | 1.00 | 0.154 | 0.615 | 283.89 | 283.50-289.30 [39-41] | |

4    Exemplary progress

    4.1 Mechanical strength: elasticity and IHPR

Elastic modulus B and yield strength σ describe the material's capability of resisting elastic and plastic deformation, respectively, at the atomic scale. Their units are the force over the acting area, [Pa] = [F/S] = E/V, being the dimensionality of energy density. At the atomic scale, the elastic modulus and yield strength are proportional to the third order differential of the crystal potential U(r) at equilibrium (r = d) and non-equilibrium (r > d) under mechanical compression [5]:

$$\left.\begin{matrix}B\\\sigma\end{matrix}\right\} \propto \left.\frac{\partial^3 U_{cryst}(r)}{\partial r^3}\right|_{r-d=x} \propto \begin{cases} \dfrac{E}{d^3}; x=0 & (elastic\ equilibrium) \\ \dfrac{U(r)}{r^3}; x>0 & (plastic\ noequilibrium) \end{cases}$$

(4)

For plastic deformation of substance at the nanometer scale, one has to consider the size effect on the melting temperature $T_m(z)$ and the bond energy $E_z(T)$:

$$\left.\begin{matrix}E_z(T)\\T_{zm}(z)\end{matrix}\right\} = \begin{cases} \eta_{1z}\left[T_{zm}(z)-T\right]+\eta_{2z} & (T\text{-}dependent\ bond\ energy) \\ z_{ib}C_z^{-m}E_b & (CN\ resolved\ T_m) \end{cases}$$

(5)

The $\eta_{1z}$ is the specific heat of linear approximation and the $\eta_{2z}$ is the energy required for dissociating a bond between z-coordinated atoms at the temperature of melting $T_{zm}$ and above. In fact, the plasticity happens only when the $T_{zm} - T$ is sufficiently small. This situation can be realized by either raising the temperature of operation or lowering the $T_{zm}$ by atomic *CN* reduction through decreasing the feature size.

Because of the competition of energy density gain and atomic cohesive energy loss, there will be a transition between the elastic deformation and plastic deformation, which is the case of the inverse Hall-Petch relationship (IHPR), see Figure 7 inset formulation for the size dependence of the elastic modulus and plastic yield strength that involves no $\eta_{2z}$ energy.

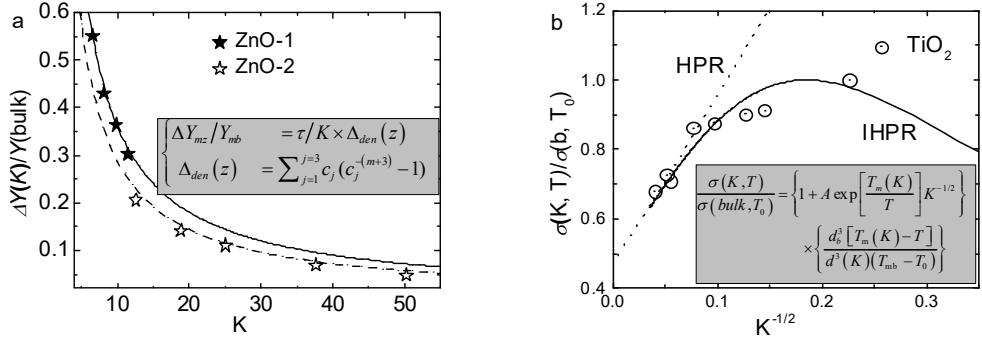

Figure 7 BOLS reproduction (solid lines) of the measured (scattered data) (a) Young's modulus enhancement for ZnO nanowires [5, 42] and (b) the IHPR (scattered datum) of (b) $TiO_2$ nanocrystals [43]. Interception at the vertical axis calibrate the bulk value of yield strength so the maximal strength is 200 % of the bulk. Insets show the respective BOLS formulations.

Figure 7 shows the theoretical reproduction of the size dependence of elastic modulus and the IHPR of the core-shell structured ZnO [5, 42] and $TiO_2$ [43] nanocrystals. In the IHPR, the thermal activation term $\exp[T_m(K)/T]$ and the energy size-dependent energy density $E(T)/d^3$ are introduced to the conventional HPR $\sigma(K) = 1+AK^{-1/2}$ to feature the size dependent yield strength of nanocrystals. With the known bulk melting temperature and bond length, we have reproduced the IHPR of a number of nanocrystals and alloys [5].

Theoretical reproduction of the mechanical strength of the skin-shell and a solid over the whole range of sizes reminds the significance of bond contraction in the surface of nanostructures. The surface is harder and more elastic than the bulk at temperatures far below $T_m$ but the surface melts more easily compared to the bulk interior of a substance. The temperature separation ($T_m - T$) plays an important role in determining the mechanical strength and ductility of a nanocrystal. The IHPR originates from the competition between the energy density gain and the atomic cohesive energy loss due to atomic undercoordination or crystal size reduction. As solid size decreases, a transition from dominance of bond strength gain to dominance of bond order loss occurs at the IHPR critical size because of the increased portion of lower coordinated atoms. During the transition, both bond order loss and bond strength gain contribute competitively.

### 4.2 Thermal stability: $T_C$ modulation and superplasticity

Figure 8 show the BOLS reproduction of the size dependent $T_m(K)$ for Au nanocrystals on different substrates [5] and the thermal strain $\varepsilon(z, T, P)$ of gold atomic chain of 5-7 atoms [5]. Insets show the respective formulation. The strain $\varepsilon(z = 2, T, P)$ depends apparently on the $T_{mz}$ -T separation and independently on the force of stretching P.

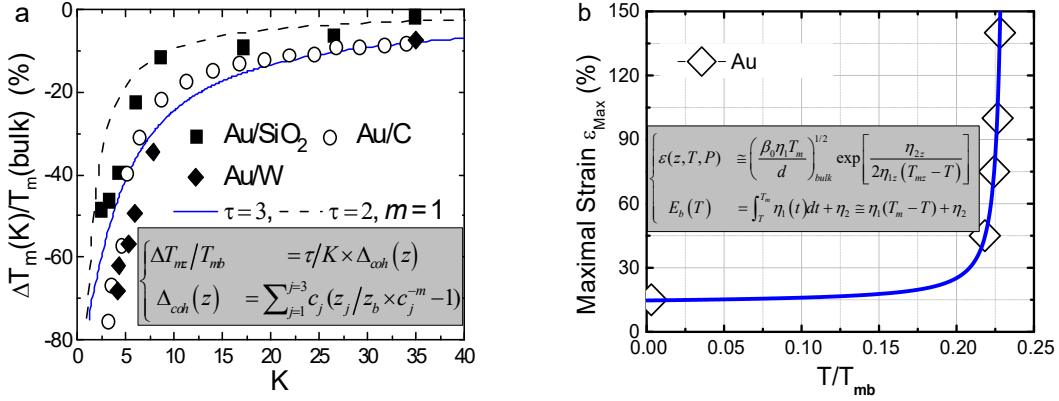

Figure 8    (a) Size dependence of the melting temperature of Au crystals and (b) temperature dependence of the strain limit of Au-Au atomic chain ($z = 2$) that melts at $0.239T_m= 318$ K. The maximal strain at 4.2 K is $0.23 \pm 0.04$ nm and reaches its limit of 60% (0.48 nm) at temperature 304 K, 12 K below its $T_m$ [5].

### 4.3 Photoelectronics: light emission and dielectric depression

Figure 9 shows theoretical reproduction of the size dependence of the bandgap of nanostructured silicon (inset shows the Si nanowire image) measured using STS conductance and optical methods. The intrinsic bandgap can be obtained directly from STS measurement or indirectly by averaging the bandgaps of photoemission and photo absorption that involve the Stokes shift (W) due to electron-phono coupling. As the bandgap is proportional to the cohesive energy per bond via the nearly-free electron approximation, the bandgap expands when the size of a semiconductor shrinks [44]. For metals, the valence band will split, generating the artificial band gap, which may explain why a conductor turns to be an insulator or a semiconductor when its size turns to be the nanoscale [45], such as Au [46] and Pd [47] nanostructures. The artificial bandgap increases with the size reduction of metallic clusters. STS conductance measurements showed that the bandgap creates and then expands to 0.7 eV when the crystalline Pd particles reduces its diameter from 4.0 to 1.6 nm [47] associated with polarization.

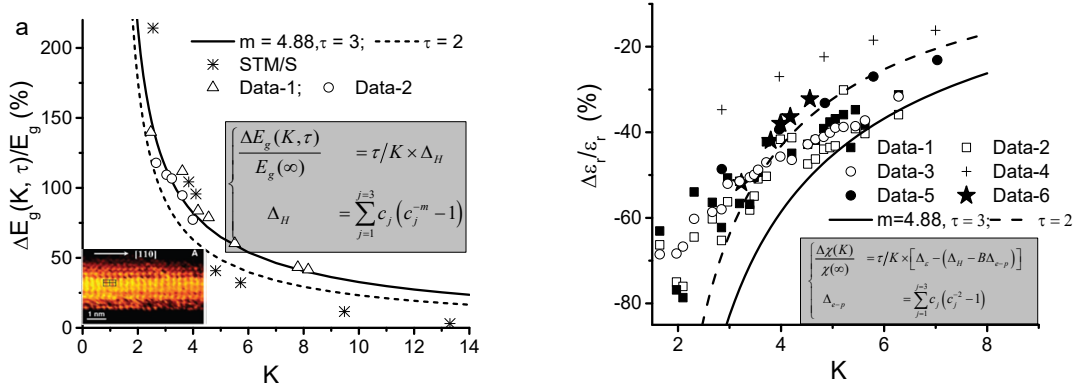

Figure 9    BOLS theoretical reproduction (solid lines) of the measured (scattered data) intrinsic size dependence of (a) the bandgap $E_g$ using STS [44] and optical method, Data –1 [48]($E_G = E_{PA} - W$), Data-2 [$E_G = (E_{PL} + E_{PA})/2$] [5], and (b) the real-part of the dielectric constant of Si [5]. Data–1, 2, 3 are after [49] Data–4 and 5 after [50] and Data–6 after [5].

With neither igniting electron-phonon interaction at 4 K temperature nor electron-hole pair (exciton) production or combination, as the vehicle for the quantum confinement theory, STS measurements revealed that [44] the $E_g$ of Si nanorods increases from 1.1 eV to 3.5 eV when the wire diameter is reduced from 7.0 to 1.3 nm. At the same time, the surface Si-Si bond contracts by ~12% from the bulk value of 0.263 to ~0.230 nm. This observation concurs excitingly with the BOLS expectation: CN-imperfection shortens and strengthens the remaining bonds of the lower-coordinated atoms associated with $E_g$ expansion that is proportional to the single bond energy. Furthermore, there is no freely moving electrons or holes in the covalently bonded semiconductors. Similarly, the size-enlarged $E_g$ of Si nanorods, Si nanodots, Ge nanostructures, and other III-V and II-VI semiconductors at the nanoscale follows closely the BOLS prediction without involving electron-hole pair interaction, electron-phonon coupling or quantum confinement [5].

The complex dielectric constant, $\varepsilon_r(\omega) = \text{Re}[\varepsilon_r(\infty)] + i\text{Im}[\varepsilon'_r(\omega)]$, is a direct measure of electron polarization response to external electric field, which has enormous impact on the electrical and optical performance of a solid and related devices. Miniaturization of a semiconductor solid to the nanometer scale often lowers the $\varepsilon_r(K)$ [50, 51]. The $\varepsilon_r(K)$ reduction enhances the Coulomb interaction between charged particles such as electrons, holes, and ionized shallow impurities in nanometric devices, leading to abnormal responses.

BOLS formulation of the static dielectric constant and the permeability, $\varepsilon_r(\infty) - 1 = \chi$, suggested that the permeability depression depends on the optical PL band shift and the lattice strain, which is expressed as follows:

$$\frac{\Delta\chi(K)}{\chi(\infty)} = -\frac{\Delta E_{PL}(K)}{E_{PL}(\infty)} - \frac{\Delta q_0(K)}{q(\infty)}$$
$$= -\frac{\Delta E_{PL}(K)}{E_{PL}(\infty)} + \frac{\Delta d_i(K)}{d_0}$$
$$= \frac{\tau}{K}\left[-(\Delta_H - B\Delta_{e-p}) + \Delta_\varepsilon\right]$$

(6)

Where $\Delta_H$, $B\Delta_{e-p}$, and $\Delta_\varepsilon$ corresponds to the atomic-undercoordination perturbation to the Hamiltonian, electron-phonon interaction and lattice strain (q is the reciprocal lattice constant).

Figure 9b shows the dielectric depression of nanostructured Si. Consistency in trends between BOLS predictions and the measured results evidences that the BOLS correlation describes adequately the true situation in which atomic CN imperfection dictates the dielectric suppression. Dielectric depression arises from the skin dominance of crystal binding ($E_g$ expansion), electron-phonon coupling, and bond contraction.

### 4.4 Chemical reactivity and atomistic catalysis

Electrons associated with point defects [25, 52], homogeneous adatoms [53], terrace edges [5, 54, 55], atomic chain ends [56, 57], and flat surfaces [58, 59] result in, for instance, new types of energy states that enhance tremendously the local catalytic reactivity at these sites even though the bulk parent, like gold, is chemically inert. Figure 10 shows that every third row of Au atoms adding to a fully Au-covered $TiO_2$ surface could improve the efficiency of CO oxidation at room temperature by a factor of 50 compared with the otherwise fully Au-covered surface [60]. The particle size-reduction could raise the reaction yield (%) of transiting 4-tert-butylbenzaldehyde to 4-tert-butylbenzoic acid by aerobic oxidation of using (a) $O_2$ and (b) air as the oxidant [61].

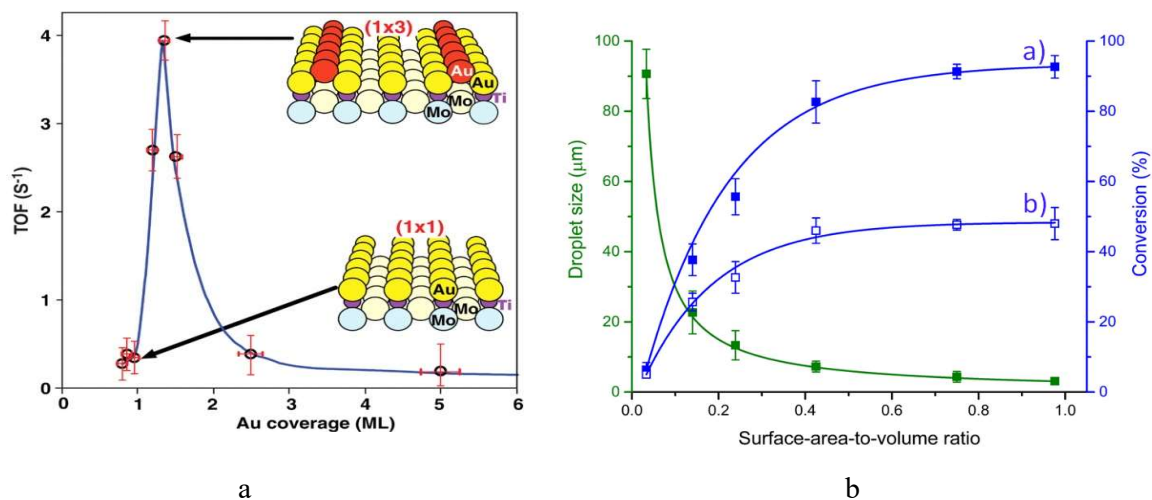

a    b

Figure 10. (a) Atomic undercoordination enhanced (a) catalytic reactivity of Au/TiO$_2$ for CO room temperature oxidation [60, 62] and (b) reaction yield (%)of the aerobic oxidation of 4-tert-butylbenzaldehyde to 4-tert-butylbenzoic acid using (a) O$_2$ and (b) air as the oxidant [61].

The activation energy for N$_2$ dissociation on Ru(0001) surface is 1.5 eV lower at the step edges than that of the flat surface, which yields at 500 K a desorption rate that is at least nine orders of magnitude higher on terraces [63]. Similar trends hold for NO decomposition on Ru(0001) surface, H$_2$ dissociation on Si(001) surface [64], and low-temperature nitridation of nano-patterned Fe skin [65]. The rough Re(11$\bar{2}$1) and Re(11$\bar{2}$0) kinks showed reactivity of three orders higher in magnitude for ammonia synthesis than that of the smooth Re(0001) surface [66]. Dispersed Ir atoms enhance greatly the reducibility of the FeO$_x$ and generate oxygen vacancies, leading to excellent performance of the Ir/FeO$_x$ single-atom catalyst [67]. The lower-coordinated edged or faceted atoms account for ~70% of the total activity of catalysts containing adatoms, atomic clusters, and nanoparticles. These kinds of atoms serve as the most active sites in reaction.

A few percent of adatoms in a specimen is sufficient to lift the reaction rate of the specimen in a catalytic process. For instances, the first methane dehydrogenation process is highly favored at the Rh-adatom site on the Rh(111) surface with respect to steps or terrace edges [68, 69]; adatoms deposited on oxides can activate the C–H bond scission [70], the acetylene ciclomerization [71], and the CO oxidation [72].

The rough Re(11$\bar{2}$1) and Re(11$\bar{2}$0) kinks have shown reactivity of three orders higher in magnitude for ammonia synthesis than that of the smooth Re(0001) surface [66]. Dispersed Ir atoms enhance greatly the reducibility of the FeO$_x$ and generate oxygen vacancies, leading to excellent performance of the Ir/FeO$_x$ single-atom catalyst [67]. An addition of a certain adsorb ate roughens Ni(210), Ir(210)[73], Rh(553) and Re(12$\bar{3}$1) [74] surfaces by faceting because of the anisotropy of surface free energy. The lower-coordinated edged or faceted atoms account for ~70% of the total activity of catalysts containing adatoms, atomic clusters, and nanoparticles. These kinds of atoms serve as the most active sites in reaction.

A few percent of adatoms in a specimen is sufficient to lift the reaction rate of the specimen in a catalytic process. For instances, the first methane dehydrogenation process is highly favored at the Rh-adatom site on the Rh(111) surface with respect to steps or terrace edges [68, 69]; adatoms deposited on oxides can activate the C–H bond scission [70], the acetylene ciclomerization [71], and the CO oxidation [72].

Besides the atomic undercoordination, lattice strain also contributes to the reactivity. Argon plasma implantation into the Ru(0001) subsurface can stretch the lattice and hence promotes adsorption of O and CO [75, 76] and enhances the NO dissociation probability on the stretched regions [77]. In the case of a supported nanoparticle catalyst, adsorption on small clusters can induce a considerable strain in the skin [78]. In any case, the existence of strain, originated by surface defects, implantation, or by interaction with the support, turns to be efficient means enhancing the surface catalytic ability [45] because of the tunable electroaffinity and work function [5].

The extremely-high catalytic efficiency of undercoordinated atoms is indeed fascinating but the fundamental nature behind remains open for exploitation. The catalytic activity of gold, for instance, was attributed to the presence of the neutral gold adatoms [79]. These adatoms differ from those at the flat surface in three ways that might enhance their catalytic activity [60]:

1) They have fewer nearest neighbors and possibly a special bonding geometry that creates more reactive orbits compared to the otherwise fully-coordinated gold atoms.
2) They exhibit quantum size effects that may alter the electronic band structure of gold nanoparticles.
3) They may undergo electronic modification by interactions with the underlying oxide that causes partial electron donation to the gold clusters.

Therefore, comprehension of the catalytic ability of the undercoordinated atoms from the perspective of local bond relaxation and the associated electron binding-energy shift, i.e., entrapment or polarization, is of importance.

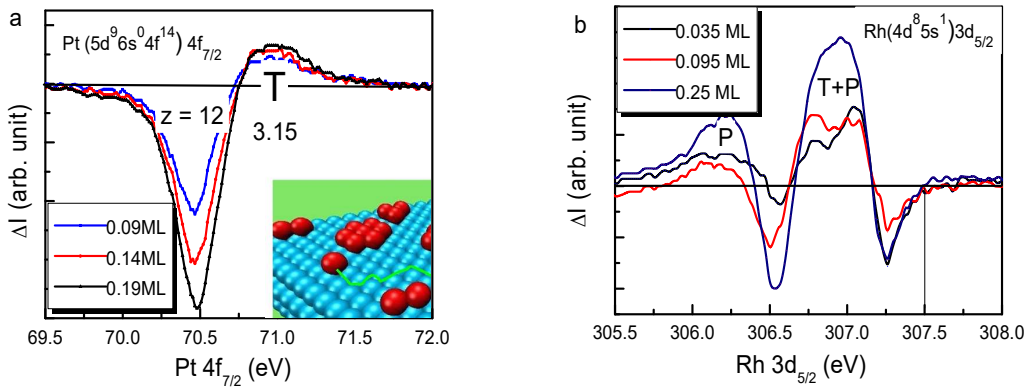

figure 11. DPS revelation of (a) Pt $4f_{7/2}$ [80] quantum entrapment (T) dominance and (b) Rh $3d_{5/2}$ [81] polarization (P) dominance as a function of adatom coverage (ML). The ZPS valley confirms the bulk component at 70.49 for Pt $4f_{7/2}$ and at 306.53 eV for Rh $3d_{5/2}$. Pt serves as an accepter-like favoring oxidation and the Rh a donor-like catalyst favoring hydrogenation [6].

In order to uncover the mechanism of the extraordinary catalytic ability, of atomic undercoordination, one can appeal to the ZPS to purify bond and electronic information due to adatoms without needing pre-specification of any spectral components. Figure 11 shows the ZPS purified energy states of homogeneous adatoms. Clearly, Pt adatoms shifts the $4f_{7/2}$ from the bulk value of 70.49 to 71.00 eV. The effective *CN* of the Pt adatoms is estimated 3.15, which is lower than the *CN* of 4.0 for an atom at the flat surface. The interatomic distance between the Pt adatoms and the Pt substrate is 17.5 % shorter and the bond is 21% stronger compared with those in the bulk.

However, the ZPS profiles for Rh adatoms are more complicated. In addition to the entrapped states at energies corresponding to $z = 4$ and 6, the polarized *P* states are present at 306.20 eV rendering the upward shift of the originally entrapped states supposed to be at $z \sim 3$. The *P* states are above the bulk valley at 306.55 eV. The valley at 307.25 eV for Rh arises from the screening and splitting of the crystal potential by adatom dipoles, which offset the entrapped states upwardly.

The absence of the *P* states in the Pt ($5d^{10}6s^04f^{14}$) $4f_{7/2}$ spectra may indicate that the empty 6s and the fully occupied $4f^{14}$ states are hardly polarizable. The difference in the *ZPS* derivatives between the Pt and the Rh adatoms coincides with the BOLS-NEP expectation that only the otherwise conductive half-filled $s$-electron Rh($4d^85s^1$) can be polarized, making the adatoms into dipoles. It is clear now why the Pt and Rh adatoms perform differently in the catalytic reaction. Pt serves as an acceptor-type being beneficial to oxidation but the Rh as a donor-type catalyst for reduction. During the reaction, Pt adatoms tend to capture electrons from the reactant while the Rh adatoms tend to donate. Along with this finding as guideline, it is possible to design and search for new catalysts at different needs using the ZPS.

4.5 Super-hydrophobicity, fluidity, lubricity, and solidity

The BOLS-NEP occurrence fosters directly phenomena of superhydrophobicity, superfluidity, superlubricity and supersolidity (called 4S for short) at the nanometer-sized contacts of liquid-solid or solid-solid. The 4S phenomena share the common characteristics of chemically non-stick, mechanically elastic, electronically repulsive, and kinetically frictionless in sliding motion or contactless when stand still [82] - a specular x-ray reflectivity analysis confirmed that an air gap of 0.3-0.5 nm thick exists between water and the hydrophobic substrate on contact [83]. What is even more fascinating is that the hydrophobic surface can switch reversibly between superhydrophobicity and superhydrophilicity when the solid surface is subject to UV radiation followed by dark aging [84, 85], as shown in Figure 12a.

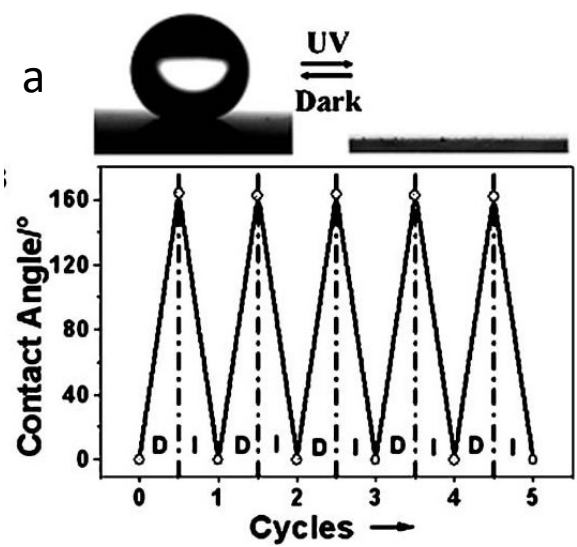

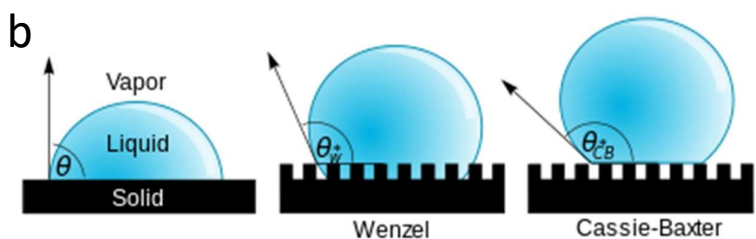

Figure 12.   (a) Reversible superhydrophobic–hydrophilic transition from 160 to 0° of the as-prepared TiO$_2$ films under the alternation of UV irradiation and dark storage [85]. (b) A droplet resting on a solid surface and surrounded by a gas forms a characteristic contact angle $\theta$. If the solid surface is rough, and the liquid is in intimate contact with the solid asperities, the droplet is in the Wenzel hydrophilic state. If the liquid rests on tops of the asperities, it is in the Cassie–Baxter hydrophobic state [86].

Many of the superhydrophobic materials found in nature display characteristics fulfilling Cassie-Baxters' law [87], which states that simply roughing up the surface can raise its contact angle. The roughening of the surface makes the hydrophobic surface even more hydrophobic and the hydrophilic surface more hydrophilic. From geometrical and mechanical point of view, a fluid can slip frictionlessly past pockets of air between textured surfaces with micrometer-scale grooves or posts of tiny distances [88], as illustrated Figure 12b.

The transport of fluid in and around nanometer-sized objects with at least one characteristic dimension below 100 nm enables the superfluidic occurrence that is impossible on larger length scales [89]. The water occupies only 60% of the cross-section area of the microchannel with air gap surrounding the fluid [90]. With absence of the air pockets, the interface between channel inner surface and the supersolid drop shows superfluidity in mass and thermal transport.

The 4S phenomena share a common elastic and repulsive origin in addition to the energetic and geometric descriptions of the existing theories. Considerations from the perspectives of surface roughness, air pockets, and surface energies are insufficient because the chemical and electronical identities do alter at the contacting skins [91]. In particular, the hydrophobicity-hydrophilicity recycling caused by UV irradiation and the subsequent dark aging is beyond the scope of Baxter-Cassie-Wenzel's descriptions.

The presently described BOLS-NEP premise provides a mechanism for the 4S in terms of chemical, electronic and phononic dynamics due to the atomic undercoordination effect. According to the BOLS-NEP scheme, the small fluidic drop can be viewed as a liquid covered with a supersolid skin that is elastic, highly charged with pinned dipoles [7]. The energy density, charge density, polarizability, and the potential trap depth are bond order dependence. The curvature increase of the nanostructured apexes enhances the BOLS-NEP effect, which results in the Baxter-Cassier's premise of roughness enhancement. The localized energy densification makes the skin stiffer and the densely- and tightly-trapped bonding charges polarize nonbonding electrons, if exist, to form dipoles locked in the skin. For the surface of quantum entrapment dominance, such as Pt adatoms, see Figure 11a, further atomic undercoordination makes the hydrophobic more hydrophobic. Therefore, Coulomb repulsion between the "electric dipoles pinned in the elastic skins at contact" and the soft phonon elasticity of the supersolid skin of liquid droplet dictate the 4S. The loss of the polarized nonbonding electrons by UV excitation and its recovery by dark aging foster the hydrophobic-hydrophilic cycling transition.

In addition, the $sp^3$-orbital hybridization of F, O, N, or C upon reacting with solid atoms generates nonbonding lone pairs or unpaired edge electrons that induce dipoles directing into the open end of a surface. The dipoles can be, however, demolished by UV radiation, thermal excitation, or excessively applied compression due to ionization or $sp$-orbit de-hybridization. Such a Coulomb repulsion between the negatively charged skins of the contacting objects not only lowers the effective contacting force and the friction but also prevents charge from being exchanged between the counterparts of the contact. Being similar to magnetic levitation, such Coulomb repulsion provides force driving the 4S.

### 4.6 Monolayer high-$T_C$ and topological edge superconductivity

The spin-resolved polarization by processes of atomic undercoordination and $sp^3$-orbital hybridization may contribute to the high-$T_C$ superconductors (HTSC) and topological insulator (TI) edge conductivity. Most HTSCs prefer the layered structure and the van der Waals gaps between layers serve as channels of charge transport. Dirac-Fermions associated with the even less coordinated edge atoms serve as the carriers transporting along the edges of the topological insulators. Figure 13 compares the angular-resolved photoelectron spectroscopy (ARPES) for the HTSC states for Bi-2212 and the $FeTe_{0.55}Se_{0.45}$ topological insulator [92]. They showed similar conductance but different $T_C$ values or coherent peak energies. The Bi-2212 has a peak at 30 meV with the $T_C$ of 91 (up to 136) K. Contrastingly, both the coherent peak energy and $T_C$ of the TI are one order lower compared with the Bi-2212, which indicates that the TI and HTSC share the common yet different extents of interactions characterized by the splitting of the peak cross the $E_F$ energy. The $T_C$ corresponds to the temperature of disappearance of the gap.

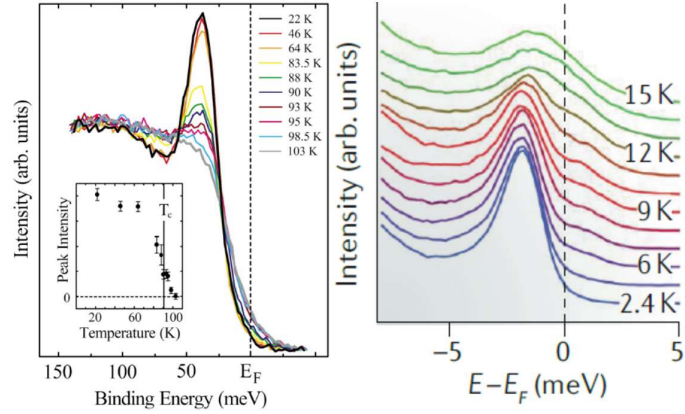

Figure 13 comparison of the (a) Bi-2212 ($Bi_2Sr_2CaCu_2O_{10+\delta}$, $Tc$ = 91 K) HTSC states [93] and the topological $FeTe_{0.55}Se_{0.45}$ [92] in the $T_C$ (12 ~ 91) K and the characteristic coherent peak energy (3 ~ 30) meV.

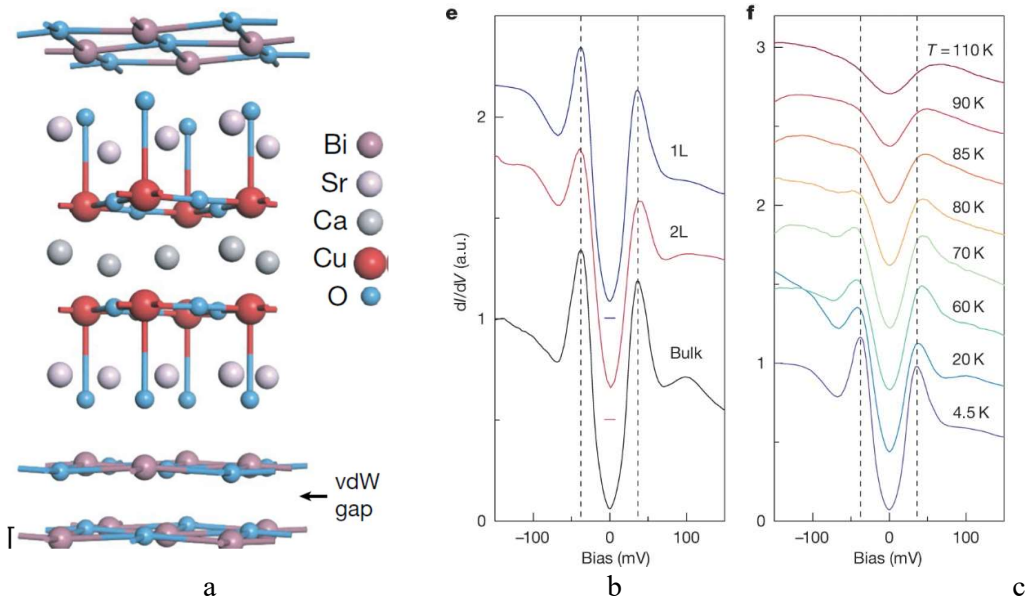

Figure 14 (a) Atomic structure of Bi-2212. The 'monolayer' refers to a half unit cell in the out-of-plane direction that contains two $CuO_2$ planes. The monolayers are separated by van der Waals gaps in bulk Bi-2212. Spatially averaged (b) differential conductance of monolayer, bilayer and bulk and temperature dependence and superconductivity transition at $T \approx 105$ K for the monolayer [94], which is identical to that for the bulk Bi-2212 at optimal doping [95].

Most strikingly, the conductivity of the HTSC Bi-2212 is thickness independence, see Figure 14. At optimal doping, the monolayer Bi-2212 has $T_C$ = 105 K [94] being identical to that of its bulk parent [95]. The monolayer refers to a half unit cell in the out-of-plane direction that contains two $CuO_2$ planes. The monolayers are separated by van der Waals gaps in bulk Bi-2212. This observation reveals not only the skin dominance of the Bi-2212 superconductivity but also the impact of the atomic undercoordination to the HTSC superconductivity.

It is even amazing, an array of tiny holes (~ 100 nm across) turns the yttrium barium copper oxide (YBCO) HTSC into a regular conductor having resistance to carrier transportation [96]. Rather than moving in concert, electron pairs conspire to stay in place, stranded on tiny islands and unable to jump to the next island, in a very thin HTSC. When the material has a current running through it and is exposed to a magnetic field, charge carriers in the YBCO will orbit the holes like water circling a drain.

The original Bardeen-Cooper-Schrieffer (BCS) theory of superconductivity adequately describes the origination and behavior of conventional superconducting metals and alloys, whose critical temperatures of superconductivity transition are below 30 K. According to the BCS theory, a large Bose-Einstein condensation resulting from the coupling of electron pairs near the Fermi surface, which are known as Cooper pairs, governs the superconductivity. Cooper pairs are named after Leon Cooper, a physics professor at Brown University who won the Nobel Prize in 1972 for describing their role in enabling superconductivity. Resistance is created when electrons rattle around in the atomic lattice of a material as they move. But when electrons join together to become Cooper pairs, they undergo a remarkable transformation. Electrons by themselves are fermions, particles that obey the Pauli exclusion principle, which means each electron tends to keep its own quantum state. Cooper pairs, however, act like bosons, which can happily share the same state. That bosonic behavior allows Cooper pairs to coordinate their movements with other sets of Cooper pairs in a way the reduces resistance to zero [97].

In 1986 and onward, copper oxides were discovered to become superconducting at temperatures up to 136 K. The emergence of the "high-$T_C$ superconductor" marked the start of a revolution in its applications as well as scientific hypotheses regarding its origin. The HTSC and the effect of atomic undercoordination in the monolayer and defect are beyond the description of the BCS theory.

One may consider first the elemental selectivity in the HTSC and TI substance. The fact that some compounds of B, C, N, O, F and elements surrounding them, particularly in group V and VI in the periodic table, form HTSC and TIs, albeit the $T_C$ and the coherent energy, implies an underlying similarity in these elements. It has been certain that N, O and F could generate nonbonding and antibonding states near the Fermi surface upon their $sp^3$-orbital hybridization. C and N can undergo $sp^2$-orbital hybridization as well with creation of the unpaired and paired electrons [98]. In turn, these localized lone pairs and the associated antibonding electrons may have a high chance of forming Cooper pairs dominating the character of the HTSCs and TIs. When an external electric field is applied, these localized pairs of electrons are easily excited and hence become highly conducting given suitable channels of transportation. Compared with the findings of graphene edge Dirac Fermin states, the effective mass of these electrons is very small and their group velocity is very high. An important characteristic is that these HTSCs all assume a two-dimensional layered structure, such as "$Cu^p : O^{-2} : Cu^p :$" chains or $CuO_2$ planes, on which superconductivity relays.

As a plausible mechanism governing the HTSC and TI, the strong correlation of electronic spins has attracted much attention. The presence of the nonbonding and the antibonding states near Fermi surface should play at least a role of competence. If the $1s$ electrons of B and C are excited to occupy the hybridized $2sp^3$ orbits, B and C would likely form valence band structures similar to those of N and O and hence result in the superconductivity. Atoms of group V and VI elements should maintain features of weak $sp$-orbital hybridization, and therefore, the nonbonding lone pairs could be a factors of dominance in both HTSC and TI conductivities. The exposition of the mechanism of HTSC from the perspective of antibond and nonbond formation and the corresponding electronics and energetics would be an approach culminating new knowledge. The spin-spin coupling may determine the coherent peak energy and thermal decoupling of the spins may correspond to the critical temperatures. From this perspective, the spin coupling in the TI is weaker than it is in the HTSC because the former has much lower $T_C$ and coherent energy and the conductivity proceeds along the even undercoordinated edge atoms.

On the other hand, atomic undercoordination enhances the polarization due to $sp$–orbital hybridization, which discriminates the skin dominance of HTSC conductivity and Dirac-Fermion generation TI edge conductivity. Because of the localization and entrapment of the polarized states, the holed HTSC transits into a regular conductor. Although it is subject to further justification, the dual process of spin-resolved nonbonding electron polarization by atomic undercoordination and *sp*-orbital hybridization may provide a feasible mechanism for the HTSC and TI superconductivity.

## 5 Summary

Exercises show consistently that the impact of atomic undercoordination to materials performance is tremendous and profound because of the undercoordination induced spontaneous bond contraction, core and bonding electron entrapment, and nonbonding electron polarization. Bond contraction raises the local density of charge and energy and the bond strength gain deepens the interatomic potential well to trap the core and bonding electrons. In turn, the locally and densely entrapped binding electrons polarize those in the valence band and above of the even-less coordinated atoms at the terminal edges and surfaces. The BOLS-NEP notion thus reconciles the unusual behaviors of undercoordinated systems and the size dependency of nanostructure on the electronic binding energy, lattice oscillating dynamics, mechanical strength, thermal stability, photon emisibility, chemical reactivity, dielectric permeability, spin-resolved insulator edge conductivity and high-$T_C$ superconductivity, etc. Consistency between theory predictions and experimental observations exemplified the validity and essentiality of the BOLS-NEP theory. Further investigation along the paths of BOLS-NEP and nonbonding electronic states would be much more challenging, fascinating, and rewarding.

Declaration

No conflicting interest is declared.

Acknowledgement

Financial Support from National Nature Science Foundation of China (No 21273191) is acknowledged.

49. M. Lannoo, C. Delerue, and G. Allan, *Screening in semiconductor nanocrystallites and its consequences for porous silicon.* Phys Rev Lett, **74**(17): 3415-3418, (1995).
50. L.W. Wang and A. Zunger, *Dielectric constants of silicon quantum dots.* Phys Rev Lett, **73**(7): 1039-1042, (1994).
51. J.P. Walter and M.L. Cohen, *Wave-vector-dependent dielectric function for Si, Ge, GaAs, and ZnSe.* Physical Review B, **2**(6): 1821-&, (1970).
52. Y. Niimi, T. Matsui, H. Kambara, and H. Fukuyama, *STM/STS measurements of two-dimensional electronic states trapped around surface defects in magnetic fields.* Physica E-low-dimensional systems & nanostructures, **34**: 100-103, (2005 ).
53. A.J. Cox, J.G. Louderback, and L.A. Bloomfield, *Experimental-observation of magnetism in Rhodium clusters.* Phys Rev Lett, **71**(6): 923-926, (1993).
54. Z. He, J. Zhou, X. Lu, and B. Corry, *Ice-like Water Structure in Carbon Nanotube (8,8) Induces Cationic Hydration Enhancement.* The Journal of Physical Chemistry C, **117**(21): 11412-11420, (2013).
55. K. Nakada, M. Fujita, G. Dresselhaus, and M.S. Dresselhaus, *Edge state in graphene ribbons: Nanometer size effect and edge shape dependence.* Phys Rev B, **54**(24): 17954-17961, (1996).
56. V.S. Stepanyuk, A.N. Klavsyuk, L. Niebergall, and P. Bruno, *End electronic states in Cu chains on Cu(111): Ab initio calculations.* Phys Rev B, **72**(15): 153407, (2005).
57. J.N. Crain and D.T. Pierce, *End states in one-dimensional atom chains.* Science, **307**(5710): 703-706, (2005).
58. T. Fauster, C. Reuss, I.L. Shumay, M. Weinelt, F. Theilmann, and A. Goldmann, *Influence of surface morphology on surface states for Cu on Cu(111).* Phys Rev B, **61**(23): 16168-16173, (2000).
59. T. Eguchi, A. Kamoshida, M. Ono, M. Hamada, R. Shoda, T. Nishio, A. Harasawa, T. Okuda, T. Kinoshita, and Y. Hasegawa, *Surface states of a Pd monolayer formed on a Au(111) surface studied by angle-resolved photoemission spectroscopy.* Phys Rev B, **74**(7): 073406, (2006).
60. M.S. Chen and D.W. Goodman, *The structure of catalytically active gold on titania.* Science, **306**(5694): 252-255, (2004).
61. X. Yan, Y.-H. Lai, and R.N. Zare, *Preparative microdroplet synthesis of carboxylic acids from aerobic oxidation of aldehydes.* Chemical science, **9**(23): 5207-5211, (2018).
62. N. Lopez, T. Janssens, B. Clausen, Y. Xu, M. Mavrikakis, T. Bligaard, and J.K. Nørskov, *On the origin of the catalytic activity of gold nanoparticles for low-temperature CO oxidation.* J Catal, **223**(1): 232-235, (2004).
63. S. Dahl, A. Logadottir, R.C. Egeberg, J.H. Larsen, I. Chorkendorff, E. Tornqvist, and J.K. Norskov, *Role of steps in $N_2$ activation on Ru(0001).* Phys Rev Lett, **83**(9): 1814-1817, (1999).
64. J. Woisetschlager, K. Gatterer, and E.C. Fuchs, *Experiments in a floating water bridge.* Exp Fluids, **48**(1): 121-131, (2010).
65. W.P. Tong, N.R. Tao, Z.B. Wang, J. Lu, and K. Lu, *Nitriding iron at lower temperatures.* Science, **299**(5607): 686-688, (2003).
66. M. Asscher and G.A. Somorjai, *The remarkable surface-structure sensitivity of the ammonia-synthesis over rhenium single-crystals.* Surface Science, **143**(1): L389-L392, (1984).
67. J. Lin, A. Wang, B. Qiao, X. Liu, X. Yang, X. Wang, J. Liang, J. Li, J. Liu, and T. Zhang, *Remarkable Performance of Ir1/FeOx Single-Atom Catalyst in Water Gas Shift Reaction.* J Am Chem Soc, **135**(41): 15314-15317, (2013).
68. A. Kokalj, N. Bonini, C. Sbraccia, S. de Gironcoli, and S. Baroni, *Engineering the reactivity of metal catalysts: A model study of methane dehydrogenation on Rh(111).* J Am Chem Soc, **126**(51): 16732-16733, (2004).
69. G. Fratesi and S. de Gironcoli, *Analysis of methane-to-methanol conversion on clean and defective Rh surfaces.* J Chem Phys, **125**(4): 044701, (2006).
70. S. Abbet, A. Sanchez, U. Heiz, W.D. Schneider, A.M. Ferrari, G. Pacchioni, and N. Rosch, *Acetylene cyclotrimerization on supported size-selected $Pd_n$ clusters (1 <= n <= 30): one atom is enough!* J Am Chem Soc, **122**(14): 3453-3457, (2000).
71. S. Abbet, U. Heiz, H. Hakkinen, and U. Landman, *CO oxidation on a single Pd atom supported on magnesia.* Phys Rev Lett, **86**(26): 5950-5953, (2001).
31